
%
%

\font\twelverm=cmr10 scaled 1200    \font\twelvei=cmmi10 scaled 1200
\font\twelvesy=cmsy10 scaled 1200   \font\twelveex=cmex10 scaled 1200
\font\twelvebf=cmbx10 scaled 1200   \font\twelvesl=cmsl10 scaled 1200
\font\twelvett=cmtt10 scaled 1200   \font\twelveit=cmti10 scaled 1200
\font\twelvesc=cmcsc10 scaled 1200  
\skewchar\twelvei='177   \skewchar\twelvesy='60


\def\twelvepoint{\normalbaselineskip=12.4pt plus 0.1pt minus 0.1pt
  \abovedisplayskip 12.4pt plus 3pt minus 9pt
  \belowdisplayskip 12.4pt plus 3pt minus 9pt
  \abovedisplayshortskip 0pt plus 3pt
  \belowdisplayshortskip 7.2pt plus 3pt minus 4pt
  \smallskipamount=3.6pt plus1.2pt minus1.2pt
  \medskipamount=7.2pt plus2.4pt minus2.4pt
  \bigskipamount=14.4pt plus4.8pt minus4.8pt
  \def\rm{\fam0\twelverm}          \def\it{\fam\itfam\twelveit}%
  \def\sl{\fam\slfam\twelvesl}     \def\bf{\fam\bffam\twelvebf}%
  \def\mit{\fam 1}                 \def\cal{\fam 2}%
  \def\sc{\twelvesc}               \def\tt{\twelvett}
  \def\sf{\twelvesf}
  \textfont0=\twelverm   \scriptfont0=\tenrm   \scriptscriptfont0=\sevenrm
  \textfont1=\twelvei    \scriptfont1=\teni    \scriptscriptfont1=\seveni
  \textfont2=\twelvesy   \scriptfont2=\tensy   \scriptscriptfont2=\sevensy
  \textfont3=\twelveex   \scriptfont3=\twelveex  \scriptscriptfont3=\twelveex
  \textfont\itfam=\twelveit
  \textfont\slfam=\twelvesl
  \textfont\bffam=\twelvebf \scriptfont\bffam=\tenbf
  \scriptscriptfont\bffam=\sevenbf
  \normalbaselines\rm}



\def\beginlinemode{\endmode
  \begingroup\parskip=0pt \obeylines\def\\{\par}\def\endmode{\par\endgroup}}
\def\beginparmode{\endmode
  \begingroup \def\endmode{\par\endgroup}}
\let\endmode=\par
{\obeylines\gdef\
{}}
\def\singlespace{\baselineskip=\normalbaselineskip}

\def\oneandahalfspace{\baselineskip=\normalbaselineskip
  \multiply\baselineskip by 3 \divide\baselineskip by 2}
\def\doublespace{\baselineskip=\normalbaselineskip \multiply\baselineskip by 2}

\newcount\firstpageno
\firstpageno=2
\footline={\ifnum\pageno<\firstpageno{\hfil}\else{\hfil\twelverm\folio\hfil}\fi}
\def\toppageno{\global\footline={\hfil}\global\headline
  ={\ifnum\pageno<\firstpageno{\hfil}\else{\hfil\twelverm\folio\hfil}\fi}}
\let\rawfootnote=\footnote              
\def\footnote#1#2{{\rm\singlespace\parindent=0pt\parskip=0pt
  \rawfootnote{#1}{#2\hfill\vrule height 0pt depth 6pt width 0pt}}}
\def\raggedcenter{\leftskip=4em plus 12em \rightskip=\leftskip
  \parindent=0pt \parfillskip=0pt \spaceskip=.3333em \xspaceskip=.5em
  \pretolerance=9999 \tolerance=9999
  \hyphenpenalty=9999 \exhyphenpenalty=9999 }
\def\dateline{\rightline{\ifcase\month\or
  January\or February\or March\or April\or May\or June\or
  July\or August\or September\or October\or November\or December\fi
  \space\number\year}}
\def\received{\vskip 3pt plus 0.2fill
 \centerline{\sl (Received\space\ifcase\month\or
  January\or February\or March\or April\or May\or June\or
  July\or August\or September\or October\or November\or December\fi
  \qquad, \number\year)}}


\hsize=6.5truein
\vsize=8.5truein  
\voffset=-1.0truein
\parskip=\medskipamount
\def\\{\cr}
\twelvepoint            
\doublespace            
\overfullrule=0pt       

\def\title                      
  {\null\vskip 3pt plus 0.2fill
   \beginlinemode \doublespace \raggedcenter \bf}

\def\author                     
  {\vskip 3pt plus 0.2fill \beginlinemode
   \singlespace \raggedcenter\sc}

\def\affil                      
  {\vskip 3pt plus 0.1fill \beginlinemode
   \oneandahalfspace \raggedcenter \sl}

\def\abstract                   
  {\vskip 3pt plus 0.3fill \beginparmode
   \singlespace ABSTRACT: }

\def\endtopmatter               
  {\endpage                     
   \body}

\def\body                       
  {\beginparmode}               

\def\head#1{                    
  \goodbreak\vskip 0.5truein    
  {\immediate\write16{#1}
   \raggedcenter \uppercase{#1}\par}
   \nobreak\vskip 0.25truein\nobreak}

\def\beginitems{
\par\medskip\bgroup\def\i##1 {\item{##1}}\def\ii##1 {\itemitem{##1}}
\leftskip=36pt\parskip=0pt}
\def\enditems{\par\egroup}

\def\beneathrel#1\under#2{\mathrel{\mathop{#2}\limits_{#1}}}

\def\refto#1{$^{#1}$}           

\def\references                 
  {\head{References}            
   \beginparmode
   \frenchspacing \parindent=0pt \leftskip=1truecm
   \parskip=8pt plus 3pt \everypar{\hangindent=\parindent}}

\gdef\refis#1{\item{#1.\ }}                     

\gdef\journal#1, #2, #3, 1#4#5#6{               
    {\sl #1~}{\bf #2}, #3 (1#4#5#6)}            

\gdef\refa#1, #2, #3, #4, 1#5#6#7.{\noindent#1, #2 {\bf #3}, #4 (1#5#6#7).\rm}

\gdef\refb#1, #2, #3, #4, 1#5#6#7.{\noindent#1 (1#5#6#7), #2 {\bf #3}, #4.\rm}

\def\pr{\journal Phys.Rev., }

\def\prl{\journal Phys.Rev.Lett., }

\def\jmp{\journal J.Math.Phys., }

\def\cmp{\journal Comm.Math.Phys., }

\def\pl{\journal Phys.Lett., }

\def\endreferences{\body}

\def\endpage                    
  {\vfill\eject}

\def\endpaper                   
  {\endmode\vfill\supereject}

\def\ref#1{Ref.~#1}                     
\def\Ref#1{Ref.~#1}                     
\def\[#1]{[\cite{#1}]}
\def\cite#1{{#1}}
\def\(#1){(\call{#1})}
\def\call#1{{#1}}
\def\taghead#1{}
\def\frac#1#2{{#1 \over #2}}
\def\half{{\frac 12}}

\def\12{{1\over2}}

\catcode`@=11
\newcount\r@fcount \r@fcount=0
\newcount\r@fcurr
\immediate\newwrite\reffile
\newif\ifr@ffile\r@ffilefalse
\def\w@rnwrite#1{\ifr@ffile\immediate\write\reffile{#1}\fi\message{#1}}

\def\writer@f#1>>{}
\def\referencefile{
  \r@ffiletrue\immediate\openout\reffile=\jobname.ref%
  \def\writer@f##1>>{\ifr@ffile\immediate\write\reffile%
    {\noexpand\refis{##1} = \csname r@fnum##1\endcsname = %
     \expandafter\expandafter\expandafter\strip@t\expandafter%
     \meaning\csname r@ftext\csname r@fnum##1\endcsname\endcsname}\fi}%
  \def\strip@t##1>>{}}

\def\citeall#1{\xdef#1##1{#1{\noexpand\cite{##1}}}}
\def\cite#1{\each@rg\citer@nge{#1}}	

\def\each@rg#1#2{{\let\thecsname=#1\expandafter\first@rg#2,\end,}}
\def\first@rg#1,{\thecsname{#1}\apply@rg}	
\def\apply@rg#1,{\ifx\end#1\let\next=\relax
\else,\thecsname{#1}\let\next=\apply@rg\fi\next}

\def\citer@nge#1{\citedor@nge#1-\end-}	
\def\citer@ngeat#1\end-{#1}
\def\citedor@nge#1-#2-{\ifx\end#2\r@featspace#1 
  \else\citel@@p{#1}{#2}\citer@ngeat\fi}	
\def\citel@@p#1#2{\ifnum#1>#2{\errmessage{Reference range #1-#2\space is bad.}%
    \errhelp{If you cite a series of references by the notation M-N, then M and
    N must be integers, and N must be greater than or equal to M.}}\else%
 {\count0=#1\count1=#2\advance\count1 by1\relax\expandafter\r@fcite\the\count0,
  \loop\advance\count0 by1\relax
    \ifnum\count0<\count1,\expandafter\r@fcite\the\count0,%
  \repeat}\fi}

\def\r@featspace#1#2 {\r@fcite#1#2,}	
\def\r@fcite#1,{\ifuncit@d{#1}
    \newr@f{#1}%
    \expandafter\gdef\csname r@ftext\number\r@fcount\endcsname%
                     {\message{Reference #1 to be supplied.}%
                      \writer@f#1>>#1 to be supplied.\par}%
 \fi%
 \csname r@fnum#1\endcsname}
\def\ifuncit@d#1{\expandafter\ifx\csname r@fnum#1\endcsname\relax}%
\def\newr@f#1{\global\advance\r@fcount by1%
    \expandafter\xdef\csname r@fnum#1\endcsname{\number\r@fcount}}

\let\r@fis=\refis			
\def\refis#1#2#3\par{\ifuncit@d{#1}
   \newr@f{#1}%
   \w@rnwrite{Reference #1=\number\r@fcount\space is not cited up to now.}\fi%
  \expandafter\gdef\csname r@ftext\csname r@fnum#1\endcsname\endcsname%
  {\writer@f#1>>#2#3\par}}

\def\ignoreuncited{
   \def\refis##1##2##3\par{\ifuncit@d{##1}%
    \else\expandafter\gdef\csname r@ftext\csname r@fnum##1\endcsname\endcsname%
     {\writer@f##1>>##2##3\par}\fi}}

\def\r@ferr{\endreferences\errmessage{I was expecting to see
\noexpand\endreferences before now;  I have inserted it here.}}
\let\r@ferences=\references
\def\references{\r@ferences\def\endmode{\r@ferr\par\endgroup}}

\let\endr@ferences=\endreferences
\def\endreferences{\r@fcurr=0
  {\loop\ifnum\r@fcurr<\r@fcount
    \advance\r@fcurr by 1\relax\expandafter\r@fis\expandafter{\number\r@fcurr}%
    \csname r@ftext\number\r@fcurr\endcsname%
  \repeat}\gdef\r@ferr{}\endr@ferences}


\let\r@fend=\endpaper\gdef\endpaper{\ifr@ffile
\immediate\write16{Cross References written on []\jobname.REF.}\fi\r@fend}

\catcode`@=12

\citeall\refto		
\citeall\ref		%
\citeall\Ref		%


\def\pp{{\prime\prime}}
\def\a{{\alpha}}
\def\b{{\beta}}
\def\d{{\delta}}
\def\D{{\Delta}}
\def\s{\sigma}
\def\half{{1 \over 2}}
\def\ra{{\rangle}}
\def\la{{\langle}}

\def\ih{{i \over \hbar}}

\def\E{{\cal E}}
\def\au{\underline \alpha}
\def\p{{\bar p}}

\def\u{{\hat u}}
\def\v{{\hat v}}
\def\x{{\hat x}}
\def\p{{\hat p}}
\def\bp{{\bar p}}
\def\bx{{\bar x}}
\def\ax{{ \langle x \rangle }}
\def\ap{{ \langle p \rangle }}
\def\A{{\hat A}}
\def\B{{\hat B}}
\def\C{{\hat C}}
\def\ria{{\rightarrow}}

\oneandahalfspace

\centerline{\bf Quantum State Diffusion, Density Matrix Diagonalization}
\centerline{\bf and Decoherent Histories: A Model}

\vskip 0.3in
\author Jonathan Halliwell and Andreas Zoupas
\affil
Theory Group, Blackett Laboratory
Imperial College, London SW7 2BZ
UK
\vskip 0.5in
\centerline{\rm PACS Numbers: 03.65.-w, 03.65.Bz, 05.40.+j, 42.50.-p}
\centerline {\rm Preprint IC 94-95/24. March, 1995}
\vskip 0.2in
\centerline {\rm Submitted to {\sl Physical Review D}}

\abstract
{We analyse the quantum evolution of a particle moving in a
potential in interaction with an environment of harmonic oscillators
in a thermal state, using the quantum state diffusion (QSD) picture
of Gisin and Percival. The QSD picture exploits a mathematical
connection between the usual Markovian master equation for the
evolution of the density operator and a class of stochastic
non-linear Schr\"odinger equations (Ito equations) for a pure state
$| \psi \ra $, and appears to supply a good description of
individual systems and processes. We find approximate stationary
solutions to the Ito equation (exact, for the case of quadratic
potentials).   The solutions are Gaussians, localized around a point
in phase space undergoing classical Brownian motion. We show, for
quadratic potentials, that every initial state approaches these
stationary solutions in the long time limit. We recover the density
operator corresponding to these solutions, and thus show, for this
particular model, that the QSD picture  effectively supplies a
prescription for approximately  diagonalizing the density operator
in a basis of phase space localized states. We show that the rate of
localization is related to the decoherence time, and also to the
timescale on which thermal and quantum fluctuations become
comparable. We use these results to exemplify the general connection
between the QSD picture and the decoherent histories approach to
quantum mechanics, discussed previously by Di\'osi, Gisin, Halliwell
and Percival.
}

\endtopmatter

\head{\bf 1. Introduction}

One of the basic premises of quantum theory is that the quantum
state of a genuinely closed and isolated system evolves according to the
Sch\"odinger equation. Although some systems of interest are
approximately closed and isolated, most of the systems we encounter
are not, as a
result of either purposeful intervention by measuring devices, or
unavoidable interaction with the immediate environment.
Such systems are said to be open, and are often studied in quantum
optics [\cite{Car}], quantum measurement theory [\cite{Zur2,Zur3}],
and in connection with decoherence and emergent classicality
[\cite{GeH,JoZ,Zeh}].

An open quantum system is in essence a distinguished subsystem of a
large, closed and isolated system in which there is a
natural division into subsystem and environment.
Although such
divisions of the world cannot be explicitly identified in general,
they do exist in a wide variety
of situations of both experimental and theoretical interest. For
example, in quantum optics, the distinguished subsystem is an atom
or small collection of atoms, and the environment is the
electromagnetic fields in interaction with it. We will in this paper
be primarily concerned with that paradigm of open quantum systems,
the quantum Brownian motion model, which consists of a
large particle coupled to a bath of harmonic oscillators in a
thermal state [\cite{CaL,QBM}].

If the state of the total quantum system is described by a
density operator $\rho_{total}$ then the state $\rho$ of the subsystem
is obtained by tracing it over the environment.
An evolution equation for $\rho$ (a master equation) may then be derived.
This is in principle obtained
quite simply by tracing the unitary evolution equation for
$\rho_{total}$ over the environment. In practice, this is hard to
carry out with any degree of generality, and has been carried out in
detail only in specific examples (see Ref.[\cite{HPZ}] for example).
As an alternative, one can ask
for the most general evolution equation for $\rho$ that preserves
density operator properties: hermiticity, unit trace, and
positivity. These conditions alone do not allow one to say very much
about the form of the equation, but
if one makes the additional assumption that the
evolution is Markovian, then the master equation must take the
Lindblad form [\cite{Lin}],
$$
{ d \rho \over dt} = -\ih  [H, \rho] - \half \sum_{j=1}^n \left(
\{ L_j^{\dag} L_j, \rho \} - 2 L_j \rho L_j^{\dag} \right)
\eqno(1.1)
$$
Here, $H$ is the Hamiltonian of the open system in the absence of
the environment (sometimes modified by terms depending on the $L_j$)
and the $n$ operators $L_j$ model the effects of the environment.
For example, in the quantum Brownian motion model, there is a single
non-hermitian $L$ which is a linear combination of position and
momentum operators.
The Markovian assumption is not always valid, but is known to be
a good approximation for a wide variety of interesting physical
situations,
{\it e.g.}, for the quantum Brownian motion model in a high
temperature environment.

Density operators evolving according to a master equation (not
always of precisely the above form) have been the subject of a number of
studies  concerned with decoherence and the emergence of classical
behaviour [\cite{Zeh,HPZ,PHZ,PaZ,UnZ,Zur,ZHP,Zur5}].
In particular models, it has been shown that the density
operator can become approximately diagonal in some basis (sometimes
more than one), indicating that interference between the states in
that basis  is destroyed. This suggests that one has some right to
regard the dynamical variables corresponding to the diagonalizing
basis as ``definite''. One may then ask for the
probabilities of successive values of these variables, and whether
those probabilities are peaked about approximately classical
evolution.

This approach to emergent classicality has considerable
intuitive appeal, but there at least two ways in which it could be
made more precise.

First of all, the notion of diagonality of the density operator is
too vague. In the quantum Brownian motion model, for example, one
expects both position and momentum to become reasonably definite.
The argument as to how this comes about often
goes as follows [\cite{UnZ}]: The
coupling to the environment is typically through position, and the
density operator tends to become approximately diagonal in position
very quickly. On longer timescales, the Hamiltonian part of the
evolution begins to contribute, and the basis of
diagonalization is rotated in phase space.
As a result of this interplay between the Hamiltonian and the
interaction with the environment, the density operator therefore becomes
approximately diagonal in a basis of states that are localized in
phase space, such as coherent states.
It is, however, difficult to see this precisely and with any
degree of generality. What is required is an explicit way of
exhibiting the diagonality in phase space localized states.

Second, the way in which
one attempts to see the emergence of classical behaviour for the
variables which have become definite is to consider the evolution of
states initially localized in phase space. Such states will tend to
follow approximately classical trajectories in phase space, with
spreading due to quantum and environmentally-induced fluctuations.
The emergence of approximately classical trajectories
is, however, much harder to see for {\it arbitrary} initial states,
{\it e.g.}, for superpositions of phase space localized states.
The problem, in essence, is that the density operator does not
in general correspond to the behaviour of an individual trajectory,
but to an ensemble. The density operator for an arbitrary initial
state will be very spread out in phase space, and it is not at all clear
that it corresponds to the intuitive
expectation of a statistical mixture of classical trajectories.
Clearly what would be very useful in this context is
an alternative description of open systems that could give a clearer
physical picture of the behaviour of an individual system, rather
than ensembles.

A recently developed picture of open systems that may be the
required precision tool
is the quantum state diffusion picture, introduced by Gisin
and Percival [\cite{GP1,GP2,GP3}]. In this picture,
the density operator
$\rho$ satisfying (1.1) is regarded
as a mean over a distribution of pure state
density operators,
$$
\rho = M | \psi \ra \la \psi |
\eqno(1.2)
$$
where $M$ denotes the mean (defined below),
with the pure states evolving according to
the non-linear stochastic Langevin-Ito equation,
$$
\eqalignno{
| d \psi \ra = -\ih H |\psi \ra dt & + \half \sum_j \left(
2 \la L_j^{\dag} \ra L_j - L_j^{\dag} L_j - \la L_j^{\dag} \ra
\la L_j \ra \right) | \psi \ra \ dt
\cr &
+ \sum_j \left( L_j - \la L_j \ra \right) | \psi \ra \ d \xi_j(t)
&(1.3) \cr }
$$
for the normalized state vector $| \psi \ra $.
Here, the $d \xi_j$ are independent complex differential random
variables representing a complex Wiener process. Their linear and
quadratic means are,
$$
M [ d \xi_j d \xi_k^* ] = \delta_{jk} \ dt, \quad
M[ d \xi_j d \xi_k ] = 0, \quad M [ d\xi_j ] = 0
\eqno(1.4)
$$

The master equation (1.1) is invariant under unitary transformations
of the Lindblad operator, $L_j \rightarrow \sum_k U_{jk} L_k $,
where $U_{jk}$ are the components of a unitary matrix [\cite{GP1}].
Physics therefore corresponds to the equivalence class of master
equations equivalent under these transformations. Correspondingly,
the Ito equation (1.3) is invariant under the same unitary transformations
on the $L_j$'s, supplemented by similar transformations on the noise
terms, and thus there is an equivalence class of Ito equations also.

The precise mathematical relation between (1.3) and (1.1) is that
the class of Ito equations (1.3) is equivalent to the
class of master equations (1.1).
Indeed, this connection supplies an alternative algorithm
for numerical solution [\cite{GP3}]. However, the strength of this picture is
that solutions to (1.3) supply an intuitively appealing picture
of the expected behavior of individual systems,
and have been seen to correspond very closely to
individual runs of actual laboratory experiments in quantum
optics [\cite{GKPTW}].

The connection between (1.1) and (1.3) is closely analagous to the
connection between the Fokker-Planck equation and the Langevin
equation in the classical description of Brownian motion. There, one
has two completely equivalent mathematical descriptions with very
different pictures. The Fokker-Planck equation describes an
ensemble of systems evolving deterministically, whilst the
Langevin equation describes an individual system evolving
stochastically.

The quantum state diffusion picture has much in common
mathematically with a variety of recent attempts to modify quantum
mechanics at a fundamental level [\cite{GRW,Dio,Pea,Per2}].
In such attempts, equations
of the form (1.3), or similar, are proposed. The difference between
QSD and such alternative formulations, is that QSD is regarded
as a phenomenological picture, appropriate only under certain
conditions, whereas the alternative formulation are taken to be
fundamental. Eq.(1.1) and (1.3) also arise in descriptions of
continuous measurement in standard quantum mechanics [\cite{CaM,Dio2}].
This paper is primarily concerned with the mathematical
properties of Eq.(1.3), hence the results will be applicable to all of
these situations.

Solutions to the Ito equation often have the feature that they
settle down to solutions of rather simple behaviour after a
period of time. This general pattern of behaviour is indicated by
numerical solutions [\cite{GP3}], along
with a number of localization
theorems, which show that, unlike evolution according to the master
equation, the dispersion of certain operators {\it decreases} as time
evolves [\cite{GP2,Per}].
That is, certain types of variables become more ``definite'' as time evolves.

A particularly useful example for our purposes was given by
Di\'osi [\cite{Dio3}], who considered the Ito equation (1.3) with $L=a x$ and
$H= {p^2 / 2m} $. (This is the quantum Brownian motion model for the
free particle neglecting dissipation). He showed that
there exist stationary solutions $ | \Psi_{pq} \ra $
to the Ito
equation that consist of Gaussian wave packets tightly concentrated
about a point in phase space evolving according to the stochastic
equations of classical Brownian motion.
This is a particularly appealing result. The
solutions to the Ito equation correspond very closely to
macroscopic observations of an individual particle
interacting with an environment.

Given a set of localized phase space solutions $ | \Psi_{pq} \ra $,
such as the Di\'osi solution,
the density operator may be reconstructed via (1.2).
This, it may be shown, may be written explicitly as
$$
\rho = \int dp dq \ f(p,q,t) | \Psi_{pq} \ra \la \Psi_{pq} |
\eqno(1.5)
$$
where $f(p,q,t)$ is a non-negative, normalized
solution to the Fokker-Planck equation
corresponding to the Langevin equation describing the Brownian motion of the
centre of the stationary solutions.

The crucial point, now, is that the representation (1.5) of the density
operator provides the desired improvements of the density
operator programme described above. Firstly, the stationary states
$  | \Psi_{pq} \ra $ are approximately orthogonal (for sufficiently
distinct values of their centres, $p,q$). Eq.(1.5)
therefore shows explicitly how the density operator may achieve a form
in which it is approximately diagonal in a set of phase space
localized states. Secondly, each diagonal element corresponds to an
individual classical trajectory (with noise). This means that the density
operator
might reasonably be interpreted as corresponding to a
statistical mixture of classical trajectories.

The object of this paper is demonstrate the above statements in detail, for
systems more general than the case considered by Di\'osi.
We will consider an open system consisting of a particle moving in a
potential $V(x)$, coupled to an environment described by Lindblad
operators in (1.1) which are a linear combination of position and
momentum operators. The detailed description of the model is given
in Section II.

We shall show, in Section III, that the Ito
equation (1.3) has stationary solutions consisting of Gaussian
wavepackets concentrated about points in phase space which undergo
classical Brownian motion. These solutions are exact for quadratic
$V(x)$. The solutions for general potentials $V(x)$ are approximate,
and are valid as long as the localization width is much smaller than
the length scale on which the potential varies.

We shall then show, in Section IV, that every initial state tends towards
one of the stationary solutions, for linear systems.
In Section V, we consider the rate of localization, and show that
it is related to the decoherence time, and also to the timescale on
which thermal and quantum fluctuations become comparable.

In Section VI, we construct the density operator of the form (1.5)
explicitly, and discuss its properties.

Arguably the most comprehensive and fundamental approach to the
problem of emergent classicality in quantum theory is the decoherent
histories approach [\cite{GeH,Gri,Omn,DoH}]. In fact,
in Ref.[\cite{DGHP}], it was argued that
there is a close connection between the quantum state diffusion
picture and the decoherent histories approach. In Section VII, we use
the above results to elaborate on this connection.

We summarize and conclude in Section VIII.

\head{\bf 2. The Model}

In this paper, we are concerned with systems described by a master
equation of the form (1.1) with a single non-hermitian
Lindblad operator linear in $\x$ and $\p$
$$
L = a \x + i b \p
\eqno(2.1)
$$
where $a$ and $b$ are real constants.
The unitary transformations under which the master equation is
invariant reduce to a simple phase invariance,
$ L \ria e^{i \theta} L $. What follows therefore applies also to $L$'s
of the form (2.1) multiplied by a phase. This form of $L$ is
sufficient to describe the quantum Brownian motion model (see
below), but also includes the cases in which $L$ is taken to be a
creation or annihilation operator.

The operator $H$ in
(1.1) is taken to be
$$
H = {\p^2 \over 2m} + V(\x) + c \{ \x, \p \} = H_0 + c \{ \x, \p \}
\eqno(2.2)
$$
where $c$ is a real constant.
The master equation may then be written,
$$
\dot \rho
= - \ih [ H_0 + (c - \half \hbar ab  ) \{ \x, \p \}, \rho ]
- i ab [ \x, \{ \rho , \p \} ]
- \half a^2 [ \x, [ \x, \rho ] ] - \half b^2 [ \p , [ \p , \rho] ]
\eqno(2.3a)
$$
or alternatively,
$$
\dot \rho = - \ih [ H_0 + (c + \half \hbar ab  ) \{ \x, \p \}, \rho ]
+ i ab [ \p, \{ \rho , \x \} ]
- \half a^2 [ \x, [ \x, \rho ] ] - \half b^2 [ \p , [ \p , \rho] ]
\eqno(2.3b)
$$
Hereafter, we take
$ c= \half ab \hbar $. This ensures that the
Ehrenfest type result, $ {\rm Tr}(\p \rho)
= {d \over dt} {\rm Tr} (\x \rho) $, holds.

The corresponding Ito equation is
$$
\eqalignno{
| d \psi \ra & = - \ih \left( H_0 + \half \hbar ab \{ \x, \p \}
\right) | \psi \ra dt
\cr & - \half \left( a^2 ( \x - \ax )^2 + b^2 ( \p - \ap )^2
+ 2 i ab (  \x - \ax \p ) - \hbar ab \right) | \psi \ra dt
\cr & + \left( a ( \x - \ax ) + i b ( \p - \ap ) \right) | \psi \ra
d \xi
&(2.4) \cr}
$$

We are particularly interested in the quantum Brownian motion model,
for which the Lindblad operator is as above, but with
$$
a = (2D)^{-\half}, \quad b = (2D)^{\half} { \gamma \over \hbar},
\quad c = \half \gamma
\eqno(2.5)
$$
Here, $ D = \hbar^2 / (8 m \gamma k T) $,
where $\gamma $ is the dissipation of the environment and $T$ is
its temperature.
The master equation in this particular case may then be written,
$$
\dot \rho
= - \ih [ H_0, \rho ]
- \ih \gamma [ \x, \{ \rho , \p \} ]
- { 2 M \gamma k T \over \hbar^2} [ \x, [ \x, \rho ] ]
- {\gamma \over 8 M k T } [ \p , [ \p , \rho] ]
\eqno(2.6)
$$
This does not, in fact, completely agree with the master equation
given in a number of previous papers on quantum Brownian motion.
In particular, the master equation given by Caldeira and Leggett
[\cite{CaL}]
does not involve the term $ [ \p, [ \p, \rho] ] $. This difference is
due to the fact the above master equation, by design, respects
the positivity of the density operator, whilst the Caldeira-Leggett
equation is known to violate it on short time scales [\cite{Amb}].
This difference is not important, since we expect the Markovian
approximation to hold only for high temperatures, and in this case
the extra term is negligible since its coefficient is
proportional to $T^{-1}$. (See Ref.[\cite{Dio7}] for further
discussion, and also Ref.[\cite{HPZ}] for the
derivation of exact master equations).

Some information on the behaviour of the solutions
to the Ito equation may be obtained
by computing the time evolution of the
moments of $\x$ and $\p$, and
this will be useful in the following sections. For any operator $G$,
the time evolution of its expectation value in the state
$| \psi \ra $ is given by
$$
\eqalignno{
d \la G \ra &= \la \psi | G | d \psi \ra
+ \la d \psi | G | \psi \ra + \la d \psi | G | d \psi \ra
\cr
& = \ih \la [ H, G ] \ra dt - \half \la L^{\dag} [ L,G]
+ [ G, L^{\dag} ] L \ra dt
\cr
& \quad\quad + \s (G^{\dag}, L ) d \xi + \s (L, G) d \xi^*
&(2.7) \cr }
$$
Here, following Percival [\cite{Per}], we have introduced the notation
$$
\s (B,C) = \la ( B^{\dag} - \la B \ra^* ) ( C - \la C \ra ) \ra
= \la B^{\dag} C \ra - \la B \ra^* \la C \ra
\eqno(2.8)
$$
for the correlation between two operators $B,C$ in the state
$| \psi \ra $.

Setting $G$ equal to $\p$ and $\x$ in this equation we obtain the
Langevin equations
$$
\eqalignno{
d \ax &= {\ap \over m } dt + \s (x, L) d \xi + \s (L, x ) d \xi^*
&(2.9) \cr
d \ap &= - \la V'(\x) \ra dt - 2 \hbar ab \ap dt + \s (p, L) d \xi
+ \s (L, p) d \xi^*
&(2.10) \cr }
$$
With the choice of parameters (2.5), and for quadratic potentials,
these equations describe
classical Brownian motion. For more general potentials, this is true only
if the state is sufficiently well-localized in $x$ for the approximation
$ \la V'(\x) \ra \approx V'(\la \x \ra ) $ to be valid (see below).

It is also of interest to compute the mean of
time evolution of higher
moments of $\x$ and $\p$, and these may again be computed using
(2.7).
One finds,
$$
\eqalignno{
M { d ( \Delta x)^2 \over dt }&=  { 2 R \over m} + 2 \hbar ab ( \Delta x)^2
+ 2 b^2 \left( { \hbar^2 \over 4} - R^2 \right) -2 a^2 (\Delta x)^4
&(2.11)\cr
M { d ( \Delta p)^2 \over dt } &=  -2 \left( \half \la \p V'(\x) + V'(\x) \p
\ra
- \ap \la V'(\x) \ra \right)
\cr & \quad \quad \quad \quad
 - 2 \hbar ab ( \Delta p)^2 + 2 a^2
\left( {\hbar^2 \over 4} - R^2 \right) - 2 b^2 ( \Delta p)^4
&(2.12)\cr
M { d R \over dt } &= - \left( \la \x V'(\x) \ra - \ax \la V'(\x) \ra \right)
+ { (\Delta p)^2 \over m }
\cr & \quad \quad \quad \quad
- 2 a^2 R  ( \Delta x)^2 - 2 b^2 R (\Delta p)^2
&(2.13)\cr }
$$
Here, $R$ is the symmetrized correlation between $\p$ and $\x$,
$$
R = \half \left( \s(x,p) + \s(p, x) \right) = \s(p, x) + {i
\hbar \over 2} = \s (x, p) - {i \hbar \over 2}
\eqno(2.14)
$$
Also, $(\D x)^4 $ denotes $\la (x - \la x \ra )^2 \ra^2 $, and
similary for $(\D p)^4 $.

To handle general potentials is too difficult except in special
cases, so approximations are required.
Under Schr\"odinger evolution in ordinary quantum mechanics
in a wide variety of potentials, there exist approximate
solutions consisting of localized Gaussian wave packets concentrated
about a classical path [\cite{Hag}].
These solutions are possible because a
sufficiently localized packet will only``notice'' the quadratic
approximation to the potential in the neighbourhood of the
wavepacket's centre. The solution breaks down after a period of
time, however, as a result of spreading of the wavepacket.

Similar types of solution to the Ito equation (2.4) are possible,
as we shall see in the next section. These have
the advantage that wavepackets tend to localize with time,
rather than spread. We may therefore justifiably approximate
the potential-dependent terms in (2.12) and (2.13) by their
expansions about the mean values of $x$ and $p$.

To see this more explicitly, and to assist the estimation of the
validity of the approximation,
introduce the notation,
$ \bx = \ax $, $\bp = \ap $, and then write the potential as,
$$
V(x) = V(\bx) + (x - \bx) V'(\bx) + \half ( x - \bx)^2
V^{\prime\prime}(\bx) + W(x, \bx)
\eqno(2.15)
$$
where
$$
W(x, \bx) = {1 \over 6} ( x - \bx )^3 V^{\prime \prime \prime}
(\bx ) + {1 \over 24} ( x - \bx)^4 V^{(4)} + \cdots
\eqno(2.16)
$$
Then the potential-dependent terms in (2.10), (2.12) and (2.13) become,
$$
\la V'(\x) \ra = V'(\bx) + \la W'(\x) \ra
\eqno(2.17)
$$
$$
\la \x V'(\x) \ra - \la x \ra \la V'(\x) \ra
= ( \D x)^2 V^{\prime\prime} (\bx) + \la ( x - \bx) W'(\x) \ra
\eqno(2.18)
$$
and
$$
\half \la \p V'(\x) + V'(\x) \p \ra - \la \p \ra \la V'(\x) \ra
= R V^{\prime\prime}(\bx)
+ \half \la \p W'(\x) + W'(\x) \p \ra - \la \p \ra \la W'(\x) \ra
\eqno(2.19)
$$
The quadratic appproximation to the potential will therefore be
valid when the terms involving $W$ may be neglected in the above
expressions. This will generally depend on the particular state.

Taking the first few terms in the Taylor expansion of $W$, Eq.(2.17)
for example, implies that the higher order terms may be neglected if
$$
\Bigl| V' (\bx) \Bigr| \ >> \ \half ( \D x )^2 \ \Bigl| V^{\prime
\prime \prime} (\bx) \Bigr|
\eqno(2.20)
$$
This is clearly the condition that the width of the state is much
less than the length scale on which the potential varies, as one
would intuitively expect. The higher order terms in (2.18) and
(2.19) also may be neglected if essentially the same type of
condition holds.

\head{\bf 3. Stationary Solutions to the Langevin-Ito Equation}

We now show how to find stationary solutions to the Langevin-Ito
equation, (2.4). It may be written
$$
| d \psi \ra = \u | \psi \ra dt + \v | \psi \ra d \xi
\eqno(3.1)
$$
where
$$
\eqalignno{
\u &= - \ih H + \half \hbar ab + i  ab \left( \ax \p - \ap \x
\right)
\cr & \quad\quad\quad\quad
- \half a^2 ( \x - \ax)^2 - \half b^2 ( \p - \ap )^2
&(3.2)\cr
\v &= L - \la L \ra
&(3.3)\cr}
$$
It is then convenient to rewrite the Ito equation in the exponential
form
$$
| \psi \ra + | d \psi \ra =
\exp \left( \u dt + \v d \xi \right) | \psi \ra
\eqno(3.4)
$$

The Diosi stationary solution has the feature that under time
evolution, its shape is preserved and the only things that change
are $ \la \p \ra $ and $ \la \x \ra $ (and possibly a phase)
[\cite{Dio3}].
Our approach to the search
for stationary solutions to our more general equation is to
require that the solution have this property.
We therefore look for solutions to (3.1) satisfying the condition,
$$
| \psi \ra + | d \psi \ra =
\exp \left( \ih {\x} d \ap - \ih {\p} d \ax + \ih d \phi \right)
| \psi \ra
\eqno(3.5)
$$
This is the statement that the state at time $t+ dt$ differs from the
state at time $t$ by nothing more than a phase, and a shift of $\ap$
and $\ax$ along the classical Brownian path described by (2.9), (2.10).
Clearly (3.5) will be satisfied for any states of the form
$$
| \psi \ra = \exp \left( \ih {\x} \ap - \ih {\p} \ax \right)
| \chi \ra
\eqno(3.6)
$$
where $| \chi \ra $ is an arbitrary fiducial state. These are
generalized coherent states [\cite{Kla}].

We will solve (3.4) and (3.5) by first combining them to yield

$$
\exp \left( \u dt + \v d \xi \right) | \psi \ra =
\exp \left( \ih {\x} d \ap - \ih {\p} d \ax + \ih d \phi \right)
| \psi \ra
\eqno(3.7)
$$
and later confirm that the solution satisfies (3.5).

Taking the operator on the right-hand side of (3.7)
over to the left-hand
side, and combining the exponentials using the
Baker-Campbell-Hausdorff formula, one obtains,
$$
\eqalignno{
\exp & \left(  - \ih {\x} d \ap + \ih {\p} d \ax - \ih d \phi
+  \u dt + \v d \xi
\right.
\cr
& \quad\quad\quad\quad \left.
 - { i \over 2 \hbar} [ \x, \v] \ d \ap d \xi \
+ { i \over 2 \hbar} [ \p, \v] \ d \ax d \xi \
\right) | \psi \ra = |
\psi \ra
&(3.8) \cr}
$$
Inserting the explicit expressions for $d \ap $, $d \ax$, $\u$ and
$\v$, and writing $ d \phi = \phi_0 dt + \phi_1 d \xi + \phi_1^* d
\xi^*$ (where $\phi_0$ is real),
this equation becomes
$$
\exp \left( \A dt + \B d \xi + \C d \xi^* \right) | \psi \ra =
| \psi \ra
\eqno(3.9)
$$
where
$$
\eqalignno{
\A &= \u + \ih \left( \la V'(\x) \ra + 2 \hbar ab \ap \right) \x
+ \ih { \ap \over m } \p + \half \s (L,L) - \ih \phi_0
&(3.10) \cr
\B &= \ih \left( - \s (p,L) \x + \s (x,L) \p - \phi_1 \right) + L -
\la L \ra
&(3.11) \cr
\C &= \ih \left( - \s ( L,p) \x + \s (L,x) \p - \phi_1^* \right)
&(3.12) \cr }
$$
Expanding the exponential in (3.9), it follows that the state must
obey the three equations,
$$
\eqalignno{
\A | \psi \ra &= 0
&(3.13) \cr
\B | \psi \ra &= 0
&(3.14) \cr
\C | \psi \ra &= 0
&(3.15) \cr }
$$

Eqs.(3.14) and (3.15) will be satisfied if
$$
\phi_1 = \s (x,L) \ap - \s (p,L) \ax
\eqno(3.16)
$$
and if the wave function is
$$
\la x | \psi \ra = N \exp \left( - \beta ( x - \ax )^2 + \ih \ap x \right)
\eqno(3.17)
$$
for some constant $\b$, to be determined.
The solution satisfies,
$$
\la x | \psi \ra + \la x | d \psi \ra =
N \exp \left( - \beta ( x - \ax - d \ax  )^2 + \ih (\ap + d \ap) x \right)
\eqno(3.18)
$$
This is clearly a generalized coherent state, and thus satisfies
Eq.(3.5).

An equation for $\beta$ may be obtained by inserting (3.17) in
(3.13). One obtains the purely algebraic equation
$$
4 \left( b^2 + {i \over m \hbar} \right) \hbar^2 \beta^2
+ 4  \hbar ab \beta - \left( a^2 + \ih V''(\ax) \right) = 0
\eqno(3.19)
$$
where we have neglected terms higher than quadratic in the
potential, as described in the previous section.

It is of course possible to write down the explicit solution for
$\b$, but it will generally be more useful in what follows to
proceed differently.
We have the uncertainty relation,
$$
( \D x)^2 ( \D p)^2 - R^2 \ge { \hbar^2 \over 4}
\eqno(3.20)
$$
with equality if and only if the state is of the
form (3.17) [\cite{DKM}].
Let us denote the values of the variances and correlation of the stationary
state (3.17) by $\s_x^2$, $\s_p^2$ and $R_0$. Then
$$
\s_x^2 \s_p^2 - R_0^2 = { \hbar^2 \over 4}
\eqno(3.21)
$$
and
$$
\beta = { ( 1 - 2i R_0 / \hbar) \over 4 \s_x^2 }
\eqno(3.22)
$$
Since, from (3.19), $\b$ is a constant (to the extent that the
approximation (2.20) holds)
the stationary values of the variances and
correlation must be those for which the right-hand sides of
(2.11)--(2.13) vanish. That is,
$$
\eqalignno{
{ R_0 \over m} + \hbar ab \s_x^2 + b^2 \left( { \hbar^2 \over 4} -
R_0^2 \right) - a^2 \s_x^4 &=0
&(3.23) \cr
- V^{\prime\prime}(\bx) R_0 - \hbar ab \s_p^2 + a^2 \left( { \hbar^2
\over 4} - R_0^2 \right) - b^2 \s_p^4 & = 0
&(3.24) \cr
- \s_x^2 V^{\prime\prime}( \bx ) + { \s_p^2 \over m} - 2 a^2 R_0
\s_x^2 - 2 b^2 R_0 \s_p^2 & = 0
&(3.25) \cr}
$$
These will be the most useful equations to work with in the following section.

To see the complete solution in a particular case, let
$V(x) = 0$ and $b=0$.
The solution for $\b$ is then,
$$
\beta = (1 - i)  \left( {m a^2 \over 8 \hbar} \right)^{\half}
\eqno(3.26)
$$
where we have chosen the square root so that ${\rm Re} \b > 0 $,
for normalizability of the state. It follows that
$$
\s_x^2 =  \left( {\hbar \over 2 m a^2} \right)^{\half},
\quad\quad
\s_p^2 = \left( { \hbar^2 m a^2 \over 2} \right)^{\half},
\quad\quad
R_0 = { \hbar \over 2}.
\eqno(3.27)
$$
This a close to minimal uncertainty state, since it satisfies,
$$
\s_p \s_x = { \hbar \over \sqrt{2} }
\eqno(3.28)
$$

The solution (3.26)--(3.28) is very similar to the solution obtained
by Di\'osi [\cite{Dio3}],
but differs by some simple numerical factors, {\it e.g.}, he obtained
$$
(\s_x^2)_{diosi} =  \left( {\hbar \over 4 m a^2} \right)^{\half},
\eqno(3.29)
$$
This difference is due to the fact that Di\'osi used an Ito equation with
a single real Wiener process, whereas the Wiener process used here
is complex.

The Di\'osi solution is also discussed in Ref.[\cite{GaG}].
Some stationary solutions to (1.3) for the harmonic oscillator have
also been found for by Salama and Gisin [\cite{SaG}], but their
choice of Lindblad operators differs from that used here.

Approximate stationary solutions to the Ito equation (2.4), for
general potentials, are currently being studied by Brun {\it et al.}
[\cite{BPSG}].

\head{\bf 4. A Localization Theorem}

We now show that all solutions to the Ito equation tend towards the
stationary solution in the long-time limit. The demonstration
applies primarily to the case of linear systems, but we will work
with a general potential in what follows, saving until the end the
issue of the extent to which that case is properly covered here.

We have shown that there is a two-parameter family of stationary
solutions, parametrized by their centres $ \ax $, $ \ap $. To prove
that all solutions tend to a stationary solution,
we will exploit the fact that the stationary solutions are
uniquely characterized by the statement that they are the
eigenfunctions of the operator
$$
A = \p - 2 i \hbar \b \x
\eqno(4.1)
$$
where $\b$ is the solution to Eq.(3.19). This means that
the stationary solutions are uniquely defined by the statement that
$ ( \Delta A )^2 = 0 $. We shall prove the desired result by showing that
$ (\Delta A )^2 $ tends to zero, in the mean.

A number of ``localization theorems'', showing that the dispersion
of certain operators decreases with time, in the mean, have been
proved by Gisin and Percival [\cite{GP2}] and by
Percival [\cite{Per}]. None of these
results is applicable to the present case because their assumptions
are too restrictive. They assume, for example, that the Hamiltonian
is zero (or negligible), or that the Lindblad operators commute with
the Hamiltonian. In brief, they assume that the Hamiltonian plays no
significant role. An important feature of the case considered in
this paper is that the stationary solutions are possible as a result
of a balance between the wavepacket spreading induced by the
Hamiltonian and the localizing effect of the Lindblad operators, and
hence the role of the Hamiltonian cannot be ignored.
An argument for the local stability of the stationary solution
in the free particle case with $b=0$ was given by Di\'osi
[\cite{Dio3}], but this proves nothing about arbitrary initial
states.

Returning to the case at hand, we have
$$
\eqalignno{
( \Delta A)^2 &= \s (A,A)
\cr
&= ( \Delta p )^2 + 4 \hbar^2 | \b |^2 ( \Delta x )^2
- 2 i \hbar ( \b + \b^* ) R  - \hbar^2 (\b + \b^* )
&(4.2) \cr}
$$
The rate of change of $ (\Delta A)^2 $ in the mean, $ M d ( \Delta
A)^2 $, is then easily computed from Eqs.(2.11)--(2.13).
It is convenient to write
$$
\eqalignno{
(\D x)^2 &= \s_x^2 (1 + X )
&(4.3)\cr
(\D p)^2 &= \s_p^2 (1 + Y)
&(4.4) \cr
R &= R_0 ( 1 + Z )
&(4.5) \cr }
$$
hence the stationary solution is $ X = Y = Z = 0 $. One then obtains,
$$
\eqalignno{
M { d ( \Delta A)^2 \over dt } =&
\ c_1 X + c_2 Y + c_3 Z
\cr &
- 2 a^2 \left( R_0^2 + { \hbar^2 \over 4} \right) X^2
- 2 b^2 \s_p^4 \ Y^2
- 2 R_0^2 \left( a^2 + b^2 { \s_p^2 \over \s_x^2} \right) Z^2
\cr &
+ 4 a^2 R_0^2 \ XZ
+ 4 b^2  { \s_p^2 \over \s_x^2} R_0^2 \ YZ
&(4.6) \cr}
$$
where
$$
\eqalignno{
c_1 &= - \hbar^2 a^2 + 2 \hbar ab \s_p^2 + 2 R_0 V^{\pp} (\bx)
&(4.7)
\cr
c_2 &= - 2 \hbar ab \s_p^2 - { 2 R_0 \over m} { \s_p^2 \over \s_x^2}
- \hbar^2 b^2 {\s_p^2 \over \s_x^2}
&(4.8)
\cr
c_3 &= { 2 R_0 \over m} { \s_p^2 \over \s_x^2} - 2 R_0 V^{\pp} (\bx)
&(4.9) \cr}
$$
and we have used (3.21) to simplify some of these expressions.

The coefficient $c_1$, $c_2$, $c_3$ have a number of useful
properties.
First, from Eq.(3.24), it is easily seen that
$$
c_1 = - { \hbar^2 a^2 \over 2} - 2 a^2 R_0^2 - 2 b^2 \s_p^4
\eqno(4.10)
$$
and thus $c_1<0$. Second, using Eq.(3.23),
$$
c_2 = - 2 { \s_p^2 \over \s_x^2} \left( a^2 \s_x^4 + b^2 R_0^2 -
{ \hbar^2 b^2 \over 4} \right) - { \hbar^2 b^2 } { \s_p^2 \over
\s_x^2}
\eqno(4.11)
$$
Using (3.21), twice, it then follows that
$$
c_2 = - 2 a^2 \s_x^2 \s_p^2 - 2b^2 \s_p^4 = c_1
\eqno(4.12)
$$
Third, $c_1 $ and $c_3 $ are related as follows.
{}From Eq.(3.25), $c_3 $ may be written,
$$
c_3 = 4 R_0^2 \left( a^2 + { \s_p^2 \over \s_x^2 } b^2 \right)
= - 2 { R_0^2 \over \s_x^2 \s_p^2 } c_1
(4.13)
$$
using (3.21) and (4.10). It follows that the linear terms in (4.6)
may now be written,
$$
c_1 X + c_2 Y + c_3 Z  = c_1 \left( X + Y - { 2 R_0^2 \over \s_x^2
\s_p^2 } Z \right)
\eqno(4.14)
$$

Clearly (4.6) is zero at the stationary solution, but it cannot be
negative for arbitrary $X$, $Y$ and $Z$, because of the presence of
the linear terms. However, $X$, $Y$ and $Z$
are not arbitrary but must respect the uncertainty principle
(an expression of which is Eq.(3.19), for example).
A convenient way to implement this restriction is
to note that
$$
0 \le ( \D A)^2 = \s_p^2 ( X+Y) - { 2 R_0^2 \over \s_x^2 } Z
\eqno(4.15)
$$
with equality if and only if the state is a general Gaussian,
such as the stationary solution. From (4.14), it is clear that
$$
c_1 X + c_2 Y + c_3 Z  = {c_1 \over \s_p^2} ( \D A)^2
\eqno(4.16)
$$
Since $c_1 < 0$, the linear terms are negative definite and zero
only at the stationary solution.

With some rearrangement of the quadratic terms, and using (3.21),
$$
\eqalignno{
M { d ( \Delta A)^2 \over dt } =&
{c_1 \over \s_p^2} ( \D A)^2
- {\hbar^2 a^2 \over 2} X^2
- 2 a^2 R_0^2 (X-Z)^2
\cr &
- 2 b^2 \s_p^4 \left( Y - { R_0^2 \over \s_p^2 \s_x^2 } Z \right)^2
- {\hbar^2 b^2 R_0^2 \over2 \s_x^4} Z^2
&(4.17) \cr}
$$
We therefore deduce that
$$
M { d ( \Delta A)^2 \over dt } \le 0
\eqno(4.18)
$$
with equality if and only if the solution is the stationary
solution. This completes the proof of localization.

As stated earlier, the stationary solutions to the Ito equation are
valid for general potentials as long as the localization width is
much less than the lengthscale on which the potential varies,
{\it i.e.}, as long as the approximation (2.20) holds.
This approximation becomes exact for linear systems.

We have essentially assumed the approximation (2.20) in proving
the above localization theorem.
This means that the proof is strictly valid only for
systems with quadratic potentials. It cannot be valid for general potentials
because even if there exist approximate stationary solutions
for which the neglect of the higher derivative terms of the
potential is valid, there will always be initial states for which
(2.20) is not a valid approximation and localization is therefore not
guaranteed for these states.
For general potentials, therefore, the
above proof implies localization only for a rather limited
class of initial states, {\it e.g.}, for states that are already close to
the stationary states.

Still, one intuitively expects that when approximate stationary
solutions exist for general potentials, there will be situations in
which most initial states will
tend towards one of those solutions. Consider, for example, the case
of a double well potential with minima a distance $L$ apart, and
suppose that the initial state has a width greater than $L$, where $L$ is
chosen so that the approximation (2.20) is not valid. Then one can
see from Eq.(2.11) that a very large initial width will be reduced
very rapidly, in the mean, bringing it into the regime in which
the approximation (2.20) is valid. Our localization theorem
would then apply. We hope to investigate this point further in a
future publication.

Note that the stationary solutions and the localization theorem do
not depend on the sign of $V^{\pp}(\bx)$, and therefore will be
valid for the upside-down harmonic oscillator (which is sometimes
used as a prototype for chaotic systems [\cite{ZuP}]).

\head{\bf 5. Localization Rate}

It is also possible to estimate the rate of localization.
Clearly,
$$
M { d ( \Delta A)^2 \over dt } \  \le
\ {c_1 \over \s_p^2} ( \D A)^2
\eqno(5.1)
$$
and thus localization proceeds on a timescale of order
$ \tau = \s_p^2 / | c_1| $. Using (4.12), this becomes
$$
\tau = \left( 2 a^2 \s_x^2 + 2 b^2 \s_p^2 \right)^{-1}
\eqno(5.2)
$$
In the quantum Brownian motion model for the free particle with
$b=0$, Eqs.(2.5), (3.27) imply that
$$
\tau  \sim \left( { \hbar \over \gamma k T} \right)^{\half}
\eqno(5.3)
$$
This, as noted previously, is the timescale on which thermal
fluctuations become comparable to the quantum ones [\cite{AnH,HuZ,AndH}].

The above represents the minimum rate of localization. The actual
rate can be much higher, {\it e.g.}, if $X$ is very large. Consider
again the free particle with $b=0$. Suppose,
the initial state consists of a superposition of wavepackets
a large distance $\ell$ apart. Then $ (\D x)^2 \sim \ell^2 $,
$$
( \D A)^2 \approx 4 \hbar^2 | \b |^2 (\D x)^2 \sim { \hbar^2 \ell^2 \over
\s_x^4}
\eqno(5.4)
$$
and the dominant contribution to the localization rate is the $X^2$
term,
$$
M { d ( \Delta A)^2 \over dt } \approx - 2 a^2
\left( R_0^2 + { \hbar^2 \over 4} \right) X^2
\sim - { \hbar^2 a^2 \ell^4 \over \s_x^4}
\eqno(5.5)
$$
It follows that in this case,
$$
\tau \sim { 1 \over \ell^2 a^2}
\eqno(5.6)
$$
For the quantum Brownian motion model, Eq.(2.5) then implies that
$$
\tau  \sim { \hbar^2 \over \ell^2 m \gamma k T}
\eqno(5.7)
$$
Both of the timescales (5.3) and (5.7) are typically exceedingly
small for macrosopic values of $m$, $\gamma$ and $T$.

As we shall show in detail in the next section, once the
solutions to the Ito equation have become localized, the
corresponding density operator has the form (1.5). The localization
timescale is therefore the timescale on which the density operator
approaches the form (1.5). Since the process of decoherence
of density operators is commonly associated with the approach to
approximately diagonal form, it is natural to regard the
localization timescale as essentially the same thing as
the decoherence timescale.

Note, however, that the so-called
``decoherence timescale'' is sometimes taken to be (5.7)
[\cite{Zur,Zur4,ZHP}].
What is clear
from the above is that the rate of approach to diagonal form
depends on initial state,
and that (5.7) is appropriate only for initial states
with very large $(\D x)^2 $.

The connections between the timescales of
decoherence and thermal fluctuations
has certainly been noted before [\cite{HuZ,AnH}], but what is new here is the
observation that both of these things are in turn related to the
timescale of localization in quantum state diffusion.

\head{\bf 6. Recovery of the Density Operator}

We now show how a density operator satisfying the master equation
may be recovered from the stationary solutions to the Ito equation.

Each solution to the Ito equation is in general
a functional of the noise term
$ \xi(t)$ over the entire history of the solution's evolution.
Eq.(1.2) indicates that the
density operator is formally recovered from these solutions by
averaging $ | \psi \ra \la \psi | $ over all possible histories of
the noise $\xi(t)$, and we write
$$
\rho = M  | \psi_\xi \ra \la \psi_\xi |
\eqno(6.1)
$$
A completely explicit form of this expression may be found in
Ref.[\cite{DGHP}] but it will not be needed here.

When the solutions $ | \psi_{\xi} \ra $ are the stationary
solutions, (3.17), they depend on the noise $\xi(t)$ only through
their centres, $\ax$, $\ap$, which obey the Langevin equations
(2.9), (2.10). We may therefore rewrite (6.1) as
$$
\rho = M  \ \int dp dq \ \delta ( p - \bp ) \ \delta (q - \bx)
\ | \psi_{pq} \ra \la \psi_{pq} |
\eqno(6.2)
$$
where we have again introduced the notation $ \bx = \ax $, $\bp = \ap $,
and $ | \psi_{pq} \ra $ denotes the stationary solution (3.17)
with centres $p$ and $q$.
The $\xi(t)$ dependence is now contained entirely in $\bp$ and
$\bx$, and Eq.(6.2) may be trivially rewritten,
$$
\rho = \int d p d q \ f(p,q,t) \ | \psi_{pq} \ra \la \psi_{pq} |
\eqno(6.3)
$$
where
$$
f(p,q,t) = M  \ \delta ( p - \bp ) \ \delta (q - \bx)
\eqno(6.4)
$$

The weight $f(p,q,t)$ is non-negative and satisfies
$$
\int dp dq \ f(p,q,t) = 1
\eqno(6.5)
$$
It is in fact the solution to the Fokker-Planck
equation corresponding to the Langevin equations. This Fokker-Planck
equation is readily derived as follows.
First note that
$$
f + df = M \ \delta ( p - \bp - d \bp ) \ \delta (q - \bx - d \bx )
\eqno(6.6)
$$
Now expanding the delta functions to second order, we have
$$
\eqalignno{
f + df =& M \ \left(
\d ( p - \bp ) \ \delta ( q - \bx )
- d \bx \ \d ( p - \bp ) \ \delta'( q - \bx )
- d \bp \ \d' ( p - \bp ) \ \delta ( q - \bx ) \right.
\cr
& \quad \quad
\left.  + \half d \bx^2 \ \d ( p - \bp ) \ \delta^{\prime\prime} ( q - \bx )
+ d \bp d \bx \ \d' ( p - \bp ) \ \delta' ( q - \bx ) \right.
\cr
& \quad \quad
\left. + \half \ d \bp^2 \d^{\prime\prime} ( p - \bp ) \ \delta ( q - \bx )
\right)
&(6.7) \cr }
$$
We may now use the Langevin equations for $\bx$ and $\bp$, and also
pull the derivatives outside the mean, $M$, for example,
$$
\eqalignno{
M \left( d \bx \ \d ( p - \bp ) \ \delta'( q - \bx ) \right)
&= M \left( { \bp \over m } \d ( p - \bp ) \ { \partial \over
\partial q} \delta( q - \bx ) \right) dt
\cr & = { p \over m } { \partial f \over \partial q } \ dt
&(6.8) \cr }
$$
We thus obtain the Fokker-Planck equation,
$$
\eqalignno{
{ \partial f \over \partial t } & = - {p \over m }
{\partial f \over \partial q}
+ V'(q) { \partial f \over \partial p}
+ 2 \hbar a b { \partial f \over \partial p }
\cr &
+ | \s(p,L) |^2 { \partial^2 f \over \partial p^2}
+ | \s(x,L) |^2 { \partial^2 f \over \partial q^2}
+ 2 {\rm Re} \left( \s(x,L) \s(L,p) \right)
{ \partial^2 f \over \partial p \partial q}
&(6.9) \cr }
$$
The coefficients of the second derivative terms are
$$
\eqalignno{
|\s(p,L)|^2 &= a^2 R_0^2 + b^2 \s_p^4 - \hbar ab \s_p^2 +
{ \hbar^2 a^2 \over 4 }
&(6.10) \cr
|\s(x,L)|^2 &= b^2 R_0^2 + a^2  \s_x^4 - \hbar ab \s_x^2
+ { \hbar^2 b^2 \over 4}
&(6.11) \cr
2 {\rm Re} \left( \s(x,L) \s(L,p) \right) &=
2 a^2 R_0 \s_x^2 + 2 b^2 \s_p^2 - 2 \hbar ab R_0
&(6.12) \cr }
$$

We have $ 2 \hbar ab = 2 \gamma $, and for high temperature, the
dominant term of the three second derivative terms is the first one,
which has coefficient,
$$
|\s(p,L)|^2 \ \approx \ 2 m \gamma k T
\eqno(6.13)
$$
The resulting Fokker-Planck equation is well-known [\cite{Ris}].
All solutions (for potentials for which $e^{-V/kT}$ is normalizable)
tend towards the stationary solution,
$$
f(p,q) = N \exp \left( - { p^2 \over2 m k T} - { V (q) \over k T}
\right)
\eqno(6.14)
$$
like $ e^{- \gamma t} $, where $N$ is a normalization factor.
For simplicity consider now the harmonic
oscillator case $V(q) = \half m \omega^2 q^2 $. Then the integrals
over $p$ and $q$ may be done explicitly, with the result,
$$
\rho(x,y) = \exp \left( - { | \beta |^2 \over \D } (x-y)^2
- { m \omega^2 ( \beta + \beta^*) \over 2 k T \D} (x^2 + y^2 )
\right)
\eqno(6.15)
$$
up to a normalization factor, where
$$
\D = { m \omega^2 \over 2 k T } + \beta + \beta^*
\eqno(6.16)
$$
For large temperature, this is readily shown to be a thermal state
[\cite{SGP}].
Similar results are expected to hold for the case of more general potentials.

To summarize, an initial density operator approaches
the form (6.3) on the localization timescale, {\it i.e.},
typically very quickly. On much longer timescales,
it will then relax to an equilibrium density operator, when one
exists for the system (it does not for the free particle, for example).

Note that although the above derivation of the asymptotic form (6.3)
strictly concerned pure initial states, it is readily extended to
mixed initial states by writing the initial state in a diagonal
basis,
$$
\rho_0 = \sum_n \ c_n \ |n \ra \la n |
\eqno(6.17)
$$
and then applying the above to each initial state
$ | n \ra \la n |$. One thus finds that the density operator tends
to the form (6.3), with $ f(p,q,t) $ of the form
$$
f(p,q,t) = \sum_n \ c_n \ f_n (p,q,t)
\eqno(6.18)
$$
where $f_n (p,q,t) $ is the solution to the Fokker-Planck equation
corresponding to the initial state $ | n \ra \la n | $.

As a final comment, note that {\it any} density operator may be written in
the form (6.3), for some function $f(p,q)$ -- this is a property
of the coherent states [\cite{Kla}].
What is special about the particular
function $f(p,q,t)$ derived here is that it is non-negative, and that it
obeys the Fokker-Planck equation (6.9). It may therefore reasonably
be interpreted as a phase space probability distribution.
(See Ref.[\cite{Dio8}] for related work on this point.)

\head{\bf 7. Connection with the Decoherent Histories Approach}

As shown in Ref.[\cite{DGHP}], there is a close connection between
the quantum state diffusion approach to open
systems and the decoherent histories approach.
In this section, we use the results of the previous sections to
exemplify and amplify this connection.

The primary mathematical aim of
the decoherent histories approach is to assign
probabilities to the possible histories of a closed system
[\cite{GeH,Gri,Omn,DoH,Hal}].
The approach is, however, applicable to open systems since they may be
regarded as subsystems of a large closed system.
A quantum-mechanical history is defined by an initial
state $\rho_0$ at time $t=t_0$ together with a string of projection operators
$ P_{\a_1} \cdots P_{\a_n}$ acting at times $t_1 \dots t_n$,
characterizing the possible alternatives of the system at those times.
The projections are exhaustive, $\sum_{\a} P_{\a} =1 $, and exclusive,
$ P_{\a} P_{\beta} = \delta_{\a \beta} P_{\a}$.
Due to interference, most sets of histories for a closed system cannot be
assigned probabilities. The interference between pairs of histories
in a set is measured by the so-called decoherence functional,
$$
D(\au, \au' ) = {\rm Tr} \left( P_{\a_n}(t_n) \cdots P_{\a_1}(t_1)
\rho P_{\a_1'}(t_1)  \cdots P_{\a_n'} (t_n) \right)
\eqno(7.1)
$$
where $P_{\a_k} (t_k) = e^{-\ih H t_k } P_{\a} e^{\ih H t_k} $,
$H$ is the Hamiltonian of the closed system
and $\au$ denotes the string $\a_1 \cdots \a_n $.
When
$$
D(\au,\au') \approx 0
\eqno(7.2)
$$
for all pairs $ \au \ne \au'$, inteference may be neglected,
and the set of histories is then said to be decoherent.
One may then
assign the probability $p(\au) = D(\au, \au)$ to the history,
which may be shown to
obey the sum rules of probability theory.

For a given Hamiltonian and initial state, one's initial aim is to find
those histories for which the decoherence condition is satisfied.
In general, it is satisfied only by
histories which are coarse-grained, which loosely speaking, means
that the projections at each moment of time give a less than
complete description of the system.
For open systems, a natural coarse-graining is to focus only on the
properties of the distinguished system itself, whilst ignoring the environment.
This involves using projections of the form,
$P_{\a} \otimes I^{\E} $ at each moment of time, where $P_{\a}$ is a
projection onto the distinguished subsystem and
$I^{\E}$ denotes the identity on the environment.
Assuming that the initial density operator factorizes, the
trace over the environment may be carried  out explicitly in the
decoherence functional (7.1), and, in the regime in which a Markovian
approximation holds, it then has the form
$$
D(\au, \au') =
{\rm Tr} \left(
P_{\a_n} K_{t_{n-1}}^{t_n}[P_{\a_{n-1}}\cdots
K_{t_1}^{t_2}[P_{\a_1} K_{t_0}^{t_1}[\rho_0] P_{\a'_1}]
\cdots P_{\a'_{n-1}}] P_{\a_n} \right)
\eqno(7.3)
$$
where the trace is now over the distinguished subsystem only.
The quantity $K_{t_k}^{t_{k+1}} $ is the reduced density operator propagator
associated with the master equation (1.1), $ \rho_t = K^t_0 [ \rho_0 ]$.

The results of the previous sections have provided us with some
information about the density operator propagator, and we can use
this information to establish some properties of the decoherence
functional (7.3).

We have seen
that any density operator will tend,
on a typically very short timescale, to the form (6.3), in which it
is approximately diagonal in a set of phase space localized states.
Once in that form, under further evolution its form will be
preserved and the only change will be that the function $f(p,q,t)$
will evolve according to the Fokker-Planck equation (6.9).

Take the projection operators in the decoherence
functional to be phase space projectors, of the
form
$$
P_{\a} = \int_{\Gamma_{\a}} dq dq \ | \psi_{pq} \ra \la \psi_{pq} |
\eqno(7.4)
$$
where $|\psi_{pq} \ra $ are the generalized coherent states (3.17),
and are eigenstates of the operator (4.1).
These quantities are not exact projection operators, but will be
approximate projectors if the phase space region $\Gamma_{\a}$ is
sufficiently large, and if its boundary is sufficiently smooth
[\cite{Omn}]. They have the property that
$ P_{\a} | \psi_{pq} \ra \approx | \psi_{pq} \ra $ if $p,q$
lie in the phase space cell $\Gamma_{\a}$, and
$ P_{\a} | \psi_{pq} \ra \approx 0 $ otherwise. Again this
approximation should be valid if $\Gamma_{\a}$ is sufficiently large
compared to the phase space area occupied by the generalized coherent
states (which is of order $\hbar$).

Consider the time evolution from $t_0$ to $t_1$ in the decoherence
functional. Clearly if this time interval is greater than the
localization time it follows from the results of Section 6 that
the density operator will evolve into the form
$$
K_{t_0}^{t_1} [ \rho_0] =
\int dp dq \ f(p,q,t_1)
| \psi_{pq} \ra \la \psi_{pq} |
\eqno(7.5)
$$
Because it is approximately diagonal in the coherent states,
it is easy to see that
$$
P_{\a_1} K_{t_0}^{t_1} [ \rho]  P_{\a_1'} \approx 0
\eqno(7.6)
$$
if $\a_1 \ne \a_1'$. Keeping only the diagonal terms, $\a_1=\a_1'$,
and evolving to time $t_2$, the (unnormalized) density operator
$ P_{\a_1} K_{t_0}^{t_1} [ \rho]  P_{\a_1} $
should again evolve into approximately diagonal
form, and again we get
$$
P_{\a_2} K_{t_1}^{t_2} \left[ P_{\a_1} K_{t_0}^{t_1} [ \rho]  P_{\a_1}
\right] P_{\a_2'} \approx 0
\eqno(7.7)
$$
if $\a_2 \ne \a_2' $.
Continuing in this way for the entire history, it is easy to see
that we will have approximate decoherence if the projections at each
moment of time are taken to be phase space projectors.
We have not estimated the degree of approximate decoherence (and this
tends to be rather involved in general), but we expect it
to be good if the size of the phase space cells is much larger
than $ \hbar $, and if the time between projections is longer than the
localization time.
We therefore find that localization in quantum state diffusion and
decoherence of histories in the decoherent histories approach
occur in the {\it same variables}.

This conclusion is in agreement with the general
connection between quantum state diffusion and decoherent histories
outlined in Ref.[\cite{DGHP}], but it also extends it somewhat.
There, it was argued that localization and decoherence tend to
occur in the
Lindblad operators. Here, the Lindblad operator is essentially
position, but we have actually obtained the stronger conclusion that
localization/decoherence occurs in the operator (4.1),
and hence, approximately, in both position {\it and} momentum.
(Note that the Lindblad operator has a small momentum part added,
but this is not the primary source of
momentum localization. Rather, it is the interplay between the
position part of the Lindblad operator and the Hamiltonian, as
discussed earlier).

Given approximate decoherence, we now consider the probabilities for
histories, given by the diagonal elements of the decoherence
functional.
{}From Eq.(7.5), and
from the properties of the phase space projections, it follows
that
$$
P_{\a_1} K_{t_0}^{t_1} [\rho_0] P_{\a_1} \approx
\int_{\Gamma_{\a_1}} dp_1 dq_1 \ f(p_1,q_1,t_1)
\ | \psi_{p_1 q_1} \ra \la \psi_{p_1 q_1} |
\eqno(7.9)
$$
Now consider the evolution from $t_1$ to $t_2$. We have, from
Section 6,
$$
K_{t_1}^{t_2} \left[ | \psi_{p_1 q_1} \ra \la \psi_{p_1 q_1} | \right]
= \int dp_2 dq_2 \ f(p_2,q_2,t_2| p_1,q_1,t_1 )
\ | \psi_{p_2 q_2} \ra \la \psi_{p_2 q_2} |
\eqno(7.10)
$$
where $ f(p_2,q_2,t_2| p_1,q_1,t_1 )$ is the solution to the
Fokker-Planck equation satisfying the initial condition,
$$
f(p_2,q_2,t_1| p_1,q_1,t_1 ) = \delta ( p_2 - p_1 ) \ \delta (q_2 -
q_1 )
\eqno(7.11)
$$
$ f(p_2,q_2,t_2| p_1,q_1,t_1 ) $
is therefore the Fokker-Planck propagator, {\it i.e.},
the probability of finding the particle at $p_2, q_2$ at time $t_2$,
given that it was at $p_1,q_1 $ at time $t_1$.
Assembling (7.9) and (7.10), it follows that
$$
\eqalignno{
P_{\a_2} K_{t_1}^{t_2} \left[ P_{\a_1} K_{t_0}^{t_1} [\rho_0 ]
P_{\a_1} \right] P_{\a_2} \approx &
\int_{\Gamma_{\a_2}} dp_2 dq_2 \ \int_{\Gamma_{\a_1}} dp_1 dq_1
\ f(p_2,q_2,t_2| p_1,q_1,t_1 )
\cr \times
& \ f(p_1, q_1, t_1 )
\ | \psi_{p_1 q_1} \ra \la \psi_{p_1 q_1} |
&(7.12) \cr }
$$
Continuing in this way for the entire history, one finds that
$$
\eqalignno{
p(\a_1, \cdots \a_n ) = &
\int_{\Gamma_{\a_n}} dp_n dq_n \ \cdots \ \int_{\Gamma_{\a_1}} dp_1 dq_1
\ f(p_n,q_n,t_n| p_{n-1},q_{n-1},t_{n-1} )
\cr & \times
\ \cdots \ f(p_2,q_2,t_2| p_1,q_1,t_1 )
\ f(p_1, q_1, t_1 )
&(7.13) \cr }
$$
This is the desired result. Eq.(7.13) is the probability that a
particle evolving according to the stochastic process described by
the Fokker-Planck equation (6.9) will be in the sequence of phase
space cells $\Gamma_{\a_1} \cdots \Gamma_{\a_n} $ at times
$t_1 \cdots t_n $.

This result is in agreement with the probabilities one would assign to
histories in the quantum state diffusion approach. For there,
once the solutions to the Ito equation have become localized, the
description of the motion on scales greater than the localization width
is classical Brownian motion according to the Langevin equations
(2.9), (2.10). This is equivalent to the description in terms of the
Fokker-Planck equation (6.9). We have therefore exemplified the
second part of the connection between quantum state diffusion and
decoherent histories put forward in Ref.[\cite{DGHP}]  -- that the
probabilities assigned to histories in each approach are the same.

A further claim in Ref.[\cite{DGHP}] is that the degree of localization is
related to the degree of decoherence. Although they are clearly related,
it is difficult to check this here because, as stated above,
explicit computation of the degree of approximate decoherence is quite
difficult. This point will be pursued in more detail elsewhere.

Finally, a property of the Fokker-Planck propagator associated with
Eq.(6.9) is that it is peaked about classical evolution
(with dissipation). It follows that the probability for histories
(7.13) will be most strongly peaked when the phase space cells
lie along a classical path.

\head{\bf 8. Summary and Discussion}

Our main results are as follows.

We have found stationary solutions to the Langevin-Ito equation
(2.4) which are exact for linear systems, and approximate for
non-linear systems as long as the localization width is much less
than the scale on which the potential varies. The solutions consist
of localized wave packets concentrated about a point in phase space
undergoing classical Brownian motion.

For linear systems, every initial state tends towards one of the
stationary solutions. For non-linear systems, some form of
localization is plausible, and will certainly be true in the
neighbourhood of the stationary solutions, but our investigations on
this point are inconclusive.

Localization proceeds on a timescale which is typically very short.
It is related to the timescale on which thermal and quantum
fluctuations become comparable, and also to the decoherence
timescale.

The density operator corresponding to the stationary solutions may
be reconstructed and has the form (1.5). It is therefore diagonal on
a set of phase space localized states. For linear systems
(and plausibly for many non-linear systems also) any
initial density operator approaches this form on the
localization time scale. On longer timescales, when dissipation is
present, the density operator approaches a thermal state (when it exists)
in the long-time limit, as expected on general grounds.
These results fulfil the aims set out in the Introduction,
concerning the density matrix approach to decoherence.

Our work also has some implications for the question of approximate
versus exact density matrix
diagonalization. As discussed in the Introduction, it is often held
important in the context of decoherence studies to find the basis in
which the density matrix is diagonal. This can of course always be
done, since the density operator is a hermitian operator,  but the
basis in which $\rho$ is exactly diagonal is generally non-trivial,
{\it i.e.}, it does not  usually consist of the eigenstates of a
simple operator. Furthermore, the basis consists of eigenstates of a
different operator at each moment of time.

Here, we have shown that the quantum state diffusion approach
naturally leads
to a basis in which the density matrix is {\it approximately}
diagonal. The basis states are the eigenstates of a simple operator,
the same operator at each moment of time.
There therefore appears to be
much to be gained by relaxing the condition of {\it exact}
diagonality.
Corresponding to these exactly and approximately diagonalizing
bases, there will be exactly and approximately decoherent set of
histories in the decoherent histories approach. In Section 7, we
exhibited the approximately decoherent set.

The bases of approximate and exact diagonality do not appear
to be ``close'' in any sense.
For example, for a Gaussian density operator (in the position representation),
the exactly diagonal basis consists of Hermite polynomials multiplied
by Gaussians (similar to energy eigenstates of the harmonic
oscillator) [\cite{JoZ}], whereas the approximately diagonal one consists of
phase space localized states. (See also Ref.[\cite{Joo}] for
examples of different bases in which the density matrix is diagonal).
This suggests that the corresponding exactly
decoherent set of histories is not necessarily ``close'' to the
approximately diagonal one, somewhat contrary to the expectation sometimes
expressed [\cite{DoK}] (although it is not clear whether there are other
exactly decoherent sets of histories that {\it are} close to the
approximate one).

The basis of states
picked out by the QSD approach appears to be ``natural'', in the
sense that they correspond to the trajectories that would actually
be observed in an individual experiment, whereas the exactly
diagonal basis does not, in general.
Correspondingly, the approximately decoherent set of histories may
seem to be more ``natural'' than the exactly decoherent set.
The question of whether one is any sense preferred over the other
is, however, a subtle one. It depends on
the sort of predictions one wishes to
make, and on the extent to which the simplified situation consisting of
a distinguished system coupled to an environment is really part of a
much larger universe in which there may be adaptive systems that can
measure different properties of the distinguished subsystem [\cite{GeH}].

The sum up, the model described in this paper illustrates
the connection between
the intuitive pictures and physical predictions
provided by the quantum state diffusion approach, density matrix
approaches, and the decoherent histories approach.
In our model, localization in quantum state
diffusion, diagonalization in the density matrix approach,
and decoherence of histories in the decoherent histories approach
all occur under the same conditions and are essentially the same
thing, for each is concerned with the conditions under which
``definite properties'' may be assigned to the system.
Furthermore, the probabilities assigned to histories in the
quantum state diffusion approach and the decoherent histories
approach approximately coincide.

\head{\bf Acknowledgements}

We are very grateful to Todd Brun, Lajos Di\'osi,
Barry Garraway, Nicolas Gisin, Peter Knight, Gerard Milburn, Ian Percival
and Rudiger Schack for useful discussions.

\references

\def\pr{{\sl Phys. Rev.\ }}
\def\prl{{\sl Phys. Rev. Lett.\ }}
\def\prep{{\sl Phys. Rep.\ }}
\def\jmp{{\sl J. Math. Phys.\ }}

\def\cmp{{\sl Comm. Math. Phys.\ }}

\def\pl{{\sl Phys. Lett.\ }}

\refis{Amb} V.Ambegaokar, {\sl Ber.Bunsenges.Phys.Chem.} {\bf 95},
400 (1991). Ambegaokar in turn cites a private communication from
P.Pechukas as the origin of the observation that the master equation
suffers from a problem with positivity.

\refis{AnH} C.Anastopoulos and J.J.Halliwell, ``Generalized
uncertainty relations and long time limits for quantum Brownian
motion models'', Imperial College preprint IC 93-94/53,
gr-qc/9407039 (1994).

\refis{AndH} A.Anderson and J.J.Halliwell, \pr {\bf 48}, 2753
(1993).


\refis{BPSG} T.Brun, I.C.Percival, R.Schack and N.Gisin, in
preparation.

\refis{CaL} A.O.Caldeira and A.J.Leggett, {\sl Physica} {\bf 121A},
587 (1983).

\refis{Car} H.J.Carmichael, {\it An Open Systems Approach to Quantum
Optics} (Springer-Verlag, 1993).

\refis{CaM} C.M.Caves and G.J.Milburn, {\sl Phys.Rev.} {\bf A36}, 5543 (1987).

\refis{DoK} H.F.Dowker and A.Kent, ``On the Consistent Histories
Approach to Quantum Mechanics'', DAMTP preprint 94-48, gr-qc/9412067
(1994).

\refis{Dio} L.Di\'osi, {\sl Phys.Rev.} {\bf A40}, 1165 (1989).

\refis{Dio2} L.Di\'osi, {\sl Phys.Lett.} {\bf 129A}, 419 (1988).

\refis{Dio3} L.Di\'osi, {\sl Phys.Lett} {\bf 132A}, 233 (1988).




\refis{Dio7} L.Di\'osi, {\sl Europhys.Lett.} {\bf 22}, 1 (1993);
{\sl Physica} {\bf A199}, 517 (1993).

\refis{Dio8} L.Di\'osi, {\sl Phys.Lett} {\bf A122}, 221 (1987).

\refis{DGHP} L.Di\'osi, N.Gisin, J.Halliwell and I.C.Percival,
{\sl Phys.Rev.Lett} {\bf 74}, 203 (1995).

\refis{DKM} V.V.Dodonov, E.V.Kurmyshev and V.I.Man'ko, \pl
{\bf 79A}, 150 (1980).

\refis{DoH} H.F.Dowker and J.J.Halliwell, {\sl Phys.Rev.} {\bf D46}, 1580
(1992)

\refis{GaG} D.Gatarek and N.Gisin, {\sl J.Math.Phys.} {\bf 32},
2152 (1991).

\refis{GeH} M. Gell-Mann and J. B. Hartle, in {\it Complexity, Entropy
and the Physics of Information, SFI Studies in the Sciences of Complexity},
Vol. VIII, W. Zurek (ed.) (Addison Wesley, Reading, 1990); and in
{\it Proceedings of the Third International Symposium on the Foundations of
Quantum Mechanics in the Light of New Technology}, S. Kobayashi, H. Ezawa,
Y. Murayama and S. Nomura (eds.) (Physical Society of Japan, Tokyo, 1990);
{\sl Phys.Rev.} {\bf D47}, 3345 (1993).

\refis{Gri} R.Griffiths, {\sl J.Stat.Phys.} {\bf 36}, 219 (1984).

\refis{GRW} G.C.Ghirardi, A.Rimini and T.Weber, {\sl Phys.Rev.} {\bf
D34}, 470 (1986); G.C.Ghirardi, P.Pearle amd A.Rimini,
{\sl Phys.Rev.} {\bf A42}, 78 (1990).



\refis{GP1} N. Gisin and I.C. Percival, {\sl J.Phys.} {\bf A25},
5677 (1992); see also {\sl Phys. Lett.} {\bf A167}, 315 (1992).

\refis{GP2} N. Gisin and I.C.Percival, {\sl J.Phys.} {\bf A26},
2233 (1993).

\refis{GP3} N. Gisin and I.C.  Percival,  {\sl J.Phys.}
{\bf A26}, 2245 (1993).

\refis{GKPTW} N. Gisin, P.L. Knight, I.C. Percival, R.C. Thompson
and  D.C. Wilson, {\sl J. Mod. Optics}, {\bf 40}, 1663 (1993);
B.Garraway and P.Knight, {\sl Phys.Rev.} {\bf A49},
1266 (1994);
P.Goetsch and R.Graham, {\sl Ann.Physik} {\bf 2}, 706 (1993).

\refis{Hag} G.A.Hagedorn, \cmp {\bf 71}, 77 (1980).

\refis{Hal} J.J.Halliwell, ``A Review of the Decoherent Histories
Approach to Quantum
Mechanics'', to appear in proceedings of the Baltimore conference, {\it
Fundamental Problems in Quantum Theory},
edited by D.Greenberger, gr-qc/9407040 (1994).

\refis{HPZ} B.L.Hu, J.P.Paz and Y.Zhang, \pr {\bf D45}, 2843
(1992).

\refis{HuZ} B.L.Hu and Y.Zhang, {\sl Mod.Phys.Lett.} {\bf A8},
3575 (1993).

\refis{Joo} E.Joos, {\sl Phys.Rev.} {\bf D36}, 3285 (1987).

\refis{JoZ} E.Joos and H.D.Zeh, {\sl Z.Phys.} {\bf B59}, 223 (1985).

\refis{Kla} J.R.Klauder and E.C.G.Sudarshan, {\it Fundamentals
of Quantum Optics} (Benjamin, New York, NY, 1968); J.R.Klauder and
B.S.Skagerstam, {\it Coherent States} (World Scientific, Singapore,
1985).

\refis{Lin} G.Lindblad, {\sl Comm.Math.Phys.} {\bf 48}, 119 (1976).

\refis{Omn} R.Omn\`es, {\it The Interpretation of Quantum Mechanics}
(Princeton University Press, Princeton, 1994);
{\sl Rev.Mod.Phys.} {\bf 64}, 339 (1992), and references therein.

\refis{PHZ} J.P.Paz, S.Habib and W.Zurek, \pr {\bf D47}, 488 (1993).

\refis{PaZ} J.P.Paz and W.Zurek, \pr {\bf 48}, 2728 (1993).

\refis{Pea} P.Pearle, {\sl Phys.Rev.} {\bf D13}, 857 (1976);
\pr {\bf A48}, 913 (1993).

\refis{Per} I.C.Percival,  {\sl J.Phys.} {\bf A27}, 1003 (1994).

\refis{Per2} I.Percival, ``Primary State Diffusion'', QMW preprint (1994).

\refis{QBM} G.S.Agarwal, \pr {\bf A3}, 828 (1971); \pr {\bf A4}, 739 (1971);
H.Dekker, \pr {\bf A16}, 2116 (1977); {\sl Phys.Rep.} {\bf 80}, 1 (1991);
G.W.Ford, M.Kac and P.Mazur, \jmp {\bf 6}, 504 (1965);
H.Grabert, P.Schramm, G-L. Ingold, \prep {\bf 168}, 115 (1988);
V.Hakim and V.Ambegaokar, \pr {\bf A32}, 423 (1985);
J.Schwinger, \jmp {\bf 2}, 407 (1961);
I.R.Senitzky, \pr {\bf 119}, 670 (1960).

\refis{Ris} H.Risken, {\it The Fokker-Planck Equations: Methods of
Solution and Applications}, Second Edition (Springer-Verlag,
Berlin,1989).

\refis{SaG} Y.Salama and N.Gisin, {\sl Phys. Lett.}
{\bf 181A}, 269, (1993).

\refis{SGP} For a discussion of thermal equilibrium in the quantum
state diffusion picture, see
T.P.Spiller, B.M.Garraway and I.C.Percival,
{\sl Phys.Lett.} {\bf A179}, 63 (1993).

\refis{UnZ} W. G. Unruh and W. Zurek, {\sl Phys.Rev.} {\bf D40}, 1071 (1989).

\refis{Zeh}  H.D. Zeh, {\sl Phys. Lett.} {\bf A172}, 189 (1993);
N. Gisin and I. Percival, {\sl Phys. Lett.} {\bf 175A}, 144
(1993).

\refis{Zur} W. Zurek, {\sl Physics Today} {\bf 40}, 36 (1991)

\refis{Zur2} W. Zurek, {\sl Phys.Rev.} {\bf D24}, 1516 (1981).

\refis{Zur3} W. Zurek, {\sl Phys.Rev.} {\bf D26}, 1862 (1982).

\refis{Zur4} W. Zurek, in {\it Frontiers of Nonequilibrium Statistical
Physics}, edited by G.T.Moore and M.O.Scully (Plenum, 1986).

\refis{Zur5} W. Zurek, {\sl Prog.Theor.Phys.} {\bf 89}, 281 (1993);
and in, {\it Physical Origins of Time Asymmetry}, edited by
J. J. Halliwell, J. Perez-Mercader and W. Zurek (Cambridge
University Press, Cambridge, 1994).

\refis{ZHP} W.Zurek, S.Habib and J.P.Paz, \prl {\bf 70},
1187 (1993).

\refis{ZuP} W.Zurek and J.P.Paz, ``Decoherence, Chaos and the Second
Law'', Los Alamos preprint (1994).

\endreferences

\end